\documentclass[12pt]{article}
\usepackage{amsmath}
\usepackage{amssymb}
\usepackage{amsfonts}

\numberwithin{equation}{section}

\title{
Dynamical symmetry algebra of two superintegrable two-dimensional systems}
\author{I Marquette$^1$, C Quesne$^{2}$\\ 
{\small $^1$ School of Mathematics and Physics, The University of Queensland,} \\ 
{\small Brisbane, QLD 4072, Australia}\\ 
{\small $^2$ Physique Nucl\'eaireTh\'eorique et Physique Math\'ematique,  Universit\'e Libre de Bruxelles,} \\ 
{\small Campus de la Plaine CP229, Boulevard~du Triomphe, B-1050 Brussels, Belgium}\\
{\small E-mail: i.marquette@uq.edu.au and Christiane.Quesne@ulb.be}}
\date{ }
\begin{document}
\baselineskip=22pt plus 1pt minus 1pt
\maketitle

\begin{abstract} 
A complete classification of 2D superintegrable systems on two-dimensional conformally flat spaces has been performed over the years and 58 models, divided into 12 equivalence classes, have been obtained. We will re-examine two pseudo-Hermitian quantum systems $E_{8}$ and $E_{10}$ from such a classification by a new approach based on extra sets of ladder operators. Those extra ladder operators are exploited to obtain the generating spectrum algebra and the dynamical symmetry one. We will relate the generators of the dynamical symmetry algebra to the Hamiltonian, thus demonstrating that the latter can be written in an algebraic form. We will also link them to the integrals of motion providing the superintegrability property. This demonstrates how the dynamical symmetry algebra explains the symmetries. Furthermore, we will exploit those algebraic constructions to generate extended sets of states and give the action of the ladder operators on them. We will present polynomials of the Hamiltonian and the integrals  of motion that vanish on some of those states, then demonstrating that the sets of states not only contain eigenstates, but also generalized states. Our approach provides a natural framework for such states.
\end{abstract}


\newpage

\section{Introduction}

Classical and quantum superintegrable systems on two-dimensional conformally flat spaces have been classified in series of papers \cite{kal05a,kal05b,kal05c,kal06a,kal06b,kal06c,kal07}. In total 58 superintegrable systems were obtained which can be divided into 12 equivalence classes. The quantum models were also studied from the perspective of their quadratic symmetry algebra, for which certain types of finite and infinite dimensional irreducible representations were built using one variable realizations \cite{post08,post09}. Later it was discovered how these quantum superintegrable models are all connected via certain contractions of quadratic algebras and also to orthogonal polynomials from the Askey-Wilson scheme of orthogonal polynomials \cite{post13,mil14a,mil14b}. Those algebraic methods are in some way within the scope of Hermitian Hamiltonians and the study of their eigenstates. The spaces on which those Hamiltonians were classified are two-dimensional complex spaces, so that some Hamiltonians need to be studied from a perspective of pseudo-Hermitian quantum mechanics and related methods and, in particular, searching for generalized states. This is so far an unexplored aspect of superintegrable systems. \par
%
%
It is known that quantum systems beyond Hermitian ones admit a more complicated structure of states and are physically relevant. Among pseudo-Hermitian systems, mainly one-dimensional ones have been studied \cite{bender90,bender05,bender07,mosta02a,mosta10,bender01,nana,mosta02b,mosta03}, but also  some two- and three-dimensional systems have been considered in the context of a quadratic potential \cite{ioffe,cannata10,cannata12,barda}. In a recent series of papers \cite{marquette22a, marquette22b}, it was demonstrated how certain two- and three-dimensional pseudo-Hermitian systems with a quadratic interaction admit generalized eigenstates forming Jordan blocks. A complete set of ladder operators was obtained and exploited to construct their dynamical symmetries. When taking appropriate combinations, those algebraic structures were connected to unusual realizations of the well-known gl(2) and gl(3) Lie algebras. Those results allowed to induce the states which form Jordan blocks and, interestingly, in different ways. Various properties, such as the construction of a bi-orthogonal set of functions, were obtained. This approach was extended to various anharmonic models with quartic and sextic interaction \cite{marquette21a}, where the states were also forming Jordan blocks, which can be obtained from the ladder operators. This paper will allow to further extend the scope of those works and point out their application in the context of superintegrable systems. \par
%
%
More specifically, the purpose of this paper is to study two Hamiltonians $E_8$ and $E_{10}$, connected to the generic superintegrable system on the sphere \cite{mil14b}. We will obtain for $E_8$ and $E_{10}$ a complete set of ladder operators, the dynamical symmetry algebra, and generalized states. The identification of the underlying hidden dynamical algebra and its connection with the Hamiltonian usually provide a characterization of an algebraic Hamiltonian, exact solvability (or even possibly quasi-exact solvability) for one-dimensional Hamiltonians. This plays an important role in the search of exactly-solvable models. The present paper will allow to make progress in extending such an approach to two-dimensional systems by connecting the dynamical symmetry algebra of those two-dimensional systems with the Hamiltonian and integrals of motion.  The paper intends to develop new approaches to look at states of superintegrable systems,  given that the classification is carried out over complex spaces and for which methods need to go beyond Hermiticity.\par
%
%
The paper is organized as follows. In section~2, we review the integrals of motion and the quadratic symmetry algebra of both superintegrable systems $E_8$ and $E_{10}$. In section~3, we present a construction of ladder operators, generating spectrum algebra and states for the model $E_8$. We also give different formulas for the action of ladder operators on states and reveal the generalized eigenstates. In section~4, we construct the hidden dynamical symmetry algebra of $E_8$ and connect it with the Hamiltonian and the integrals of motion. In sections~5 and 6, we carry out the same program for the model $E_{10}$. Finally, section~7 contains the conclusion.\par
%
%
\section{\boldmath Integrals of motion and quadratic symmetry algebra of $E_8$ and $E_{10}$ }

\subsection{\boldmath Superintegrable system $E_8$}

We consider the quantum system $E_8$, which was obtained in the context of the classification of superintegrable systems \cite{kal05a,kal05b,post08,post09,mil14a}. It is connected to the generic Hamiltonian on the 2-sphere S9 via Wigner-In\"on\"u type of contraction of the symmetry algebra. In the special case where the parameter $a_2$ vanishes, the Hamiltonian is given by the following expression
\begin{equation} 
  H= - 4 \partial_z \partial_{\bar{z}} + b_1 z \bar{z} - \frac{a_3}{\bar{z}^2}, \label{eq:E8Hr}
\end{equation}
in terms of the complex variables $z = x + {\rm i}y$ and $\bar{z} = x - {\rm i}y$.\par
%
%
The Hamiltonian (\ref{eq:E8Hr}) is superintegrable, which means that it allows two algebraically independent integrals of motions $L_1$ and $L_2$, given by 
\begin{equation}
  L_1 = - \partial_z^2  + \frac{b_1}{4} \bar{z}^2 ,  \label{eq:E8L1r}
\end{equation}
\begin{equation}
  L_2 =( z \partial_z - \bar{z} \partial_{\bar{z}} )^2 - \frac{a_3 z}{\bar{z}} .  \label{eq:E8L2r}
\end{equation}
The latter both commute with Hamiltonian (\ref{eq:E8Hr})
\begin{equation}
  [H,L_1]=[H,L_2] =0, 
\end{equation}
and, using an additional linearly independent integral $R$, lead to the finitely generated quadratic algebra 
\begin{align}
  [L_1,L_2]&=R, \label{eq:E8Q1} \\
  [L_1,R]&=8 L_1^2 , \label{eq:E8Q2} \\
  [L_2,R]&=-8 \{L_1,L_2\} - 16 L_1 + 2 a_3 H. \label{eq:E8Q3}
\end{align}
\par
%
%
We note the presence of the Hamiltonian in the structure constants and also the additional
relation among $R$, $L_1$, $L_2$, and $H$, given by the expression
\begin{equation}
   R^2= -8 \{L_1^2,L_2\}-\frac{176}{3} L_1^2 + a_3 L_1 H + b_1 a_3^2 . \label{eq:E8Rrel}
\end{equation}
Such a constraint is called closure relation. The integral $R$ plays an important role in the description of the Casimir invariant and the construction of some representations.\par
%
%
\subsection{\boldmath Superintegrable system $E_{10}$ }

There is another model $E_{10}$ from the classification on conformally flat spaces which can be re-examined in our approach taking into account pseudo-Hermiticity. In the special case where the parameter $a_2$ vanishes, its Hamiltonian is
\begin{equation}
  H=-4 \partial_z \partial_{\bar{z}} + b_1 \left( z \bar{z} - \frac{1}{2} \bar{z}^3 \right)- a_3 \bar{z} 
  \label{eq:E10Hr}
\end{equation}
and its integrals of motion have the explicit form
\begin{equation}
  L_1 = - \partial_z^2 + \frac{b_1}{4} \bar{z}^2 + \frac{a_3}{12} , \label{eq:E10L1r}
\end{equation}
\begin{equation}
  L_2 = \{z \partial_z - \bar{z} \partial_{\bar{z}}, \partial_z \}- \partial_{\bar{z}}^2 + \frac{b_1 }{16}
  (2 z + \bar{z}^2 ) (2 z - 3 \bar{z}^2 ) - \frac{a_3}{4} (2z + \bar{z}^2) . \label{eq:E10L2r}
\end{equation}
The integrals satisfy the commutation relations with the Hamiltonian
\begin{equation}
  [H,L_1]=[H,L_2]=0.
\end{equation}
\par
%
%
With another linearly independent integral $R$, they lead to the following quadratic algebra, given by the three commutation relations
\begin{align}
  [L_1,L_2]&=R, \label{eq:E10Q1} \\
  [L_1,R]&= - 2 b_1 L_1 + \frac{1}{6} a_3 b_1 , \label{eq:E10Q2} \\
  [L_2,R]&=24 L_1^2 + 4a_3 L_1 + 2b_1L_2. \label{eq:E10Q3}
\end{align}
In this case, the structure constants do not depend on the Hamiltonian, but we have an extra relation
\begin{equation}
  R^2= -2 b_1 \{L_1,L_2\} -16 L_1^3 -4 a_3 L_1^2 + \frac{1}{4} b_1 H^2 + \frac{1}{3} a_3b_1  L_2 
  + \frac{1}{27}( a_3^3 - 27 b_1^2 ). \label{eq:E10Rrel}
\end{equation}
\par
%
%
In both cases $E_8$ and $E_{10}$, one-variable models were presented for these quadratic symmetry algebras. They provide information on part of the spectrum connected with particular type of representations. Those methods have been widely applied to quantum superintegrable systems, in particular in two dimensions but also in higher dimensions. To highlight how the set of integrals $\{L_1,R\}$ plays an important role in those two models, even if we have three generators quadratic algebras, and in general there is no subalgebra, here in both cases there is a two-dimensional subalgebra involving those two integrals and given by (\ref{eq:E8Q2}) and (\ref{eq:E10Q2}), respectively. 
%
%
\section{\boldmath New ladder operators for the superintegrable system $E_8$ }

\setcounter{equation}{0}

In a series of papers, the idea of a complete set of ladder operators has been introduced for pseudo-Hermitian models in two and three dimensions \cite{marquette22a,marquette22b,marquette21a}. It was demonstrated how those complete sets of ladder operators allowed to induce states forming Jordan blocks in an algebraic manner. Those states correspond to multivariate polynomials in two and three variables. In this section, we will point out how the superintegrable model $E_8$ can also be studied from this algebraic perspective and how we can generate generalized eigenstates. \par
%
%
\subsection{ Construction of ladder operators and generating spectrum algebra } 

General ladder operators $A^{\pm}$ and $B^{\pm}$ can be introduced for the Hamiltonian of $E_8$ by imposing closure of their commutation relations with the Hamiltonian. This can be obtained by allowing some functions to be present \cite{marquette21a},
\begin{equation}
  A^{\pm} = \partial_z  \mp f(\bar{z}),  \label{eq:E8Apm}
\end{equation}
\begin{equation}
  B^{\pm} = \partial_{\bar{z}} \pm g(z,\bar{z}). \label{eq:E8Bpm}
\end{equation}
For the functions $f(\bar{z})$ and $g(z,\bar{z})$, we obtain
\begin{equation}
  f(\bar{z}) = \frac{1}{2} \sqrt{b_1} \bar{z},  \label{eq:E8f}
\end{equation}
\begin{equation}
  g(z,\bar{z}) =  - \frac{1}{2} \sqrt{b_1} z  + \frac{a_3}{2 \sqrt{b_1}} \frac{1}{\bar{z}^3}. \label{eq:E8g}
\end{equation}
\par
%
%
Together with the Laurent polynomial
\begin{equation}
  M=-\frac{3a_3}{\sqrt{b_1}} \frac{1}{\bar{z}^4},  \label{eq:M}
\end{equation}
those ladder operators satisfy the following commutation relations
\begin{equation}
  [H,A^{\pm}]=\pm 2 \sqrt{b_1} A^{\pm},\quad [H,B^{\pm}]=\pm 2 \sqrt{b_1} B^{\pm} \mp 2 M A^{\pm},
   \label{eq:E8lad1} 
\end{equation}
\begin{equation}
  [A^{-},A^{+}]=0,\quad [B^{-},B^{+}]=M, \label{eq:E8lad2} 
\end{equation}
\begin{equation}
  [A^{\pm},B^{\pm}]=0,\quad [A^{\pm},B^{\mp}]=\pm \sqrt{b_1}, \label{eq:E8lad3} 
\end{equation}
\begin{equation}
  [A^{-},M]=[A^{+},M]=0, \label{eq:E8ApmM}
\end{equation}
\begin{equation}
  [B^{-},M]= \frac{4}{3 a_3 b_1} M^2 (A^{-} - A^{+})^3,  \label{eq:E8BmM}
\end{equation}
\begin{equation}
  [B^{+},M]= \frac{4}{3 a_3 b_1} M^2 (A^{-} - A^{+})^3,  \label{eq:E8BpM}
\end{equation}
\begin{equation}
  [H,M]=- \frac{8}{3 a_3 b_1} M^2 (A^{-} - A^{+})^3 (A^{-} +A^{+} ), \label{eq:E8HM}
\end{equation}
where we note the important role of $M$ in the closure relations (\ref{eq:E8lad1}) and (\ref{eq:E8lad2}).\par
%
%
As seen from the commutation relations, the operators $A^{\pm}$ and $B^{\pm}$ form a complete set of ladder operators for the two-dimensional system. Their commutation relations with $M$ do not have, however, a Lie algebraic form since some quintic or sextic relations make their appearance. Let us note that there are alternative ways of writing the commutation relations with $M$, such as the following nested commutator relations
\begin{equation}
   [B^{\epsilon_1},[B^{\epsilon_2},[B^{\epsilon_3},[B^{\epsilon_4},M]]]]=-\frac{280 \sqrt{b_1}}{a_3} M^2,  \label{eq:E8BM}
\end{equation}
where $\epsilon_1=\pm$, $\epsilon_2=\pm$, $\epsilon_3=\pm$, $\epsilon_4=\pm$.\par
%
%
Equations (\ref{eq:E8BmM}), (\ref{eq:E8BpM}), and (\ref{eq:E8HM}) allow to close the ladder operator algebra, which is of a different form than that for well-known superintegrable Hermitian systems, such as the 2D harmonic oscillator, the Verrier-Evans system or the anisotropic 2D oscillator \cite{das99,ver08,rod08,kal10,mar10}. One purpose of the present paper will be to explain how the complicated structure of those commutation relations still allows to use the ladder operators to obtain dynamical symmetries and a description of the generalized states of the two-dimensional system.\par
%
%
We give here some further commutation relations between generators of the quadratic symmetry algebra and the set of ladder operators. They are part of what is called the generating spectrum algebra, 
\begin{equation}
  [L_1, A^{\pm}]=0, \quad [L_1,B^{\pm}]= \pm \sqrt{b_1} A^{\pm}, \label{eq:E8symABM1}
\end{equation}
\begin{equation}
  [L_1,M]=0, \quad [R,M]=16 M L_1. \label{eq:E8symABM2}
\end{equation}
\par
%
%
\subsection{ Construction of states and action of operators}

In order to construct the states algebraically, we first look for the state annihilated simultaneously by the lowering operators $A^{-}$ and $B^{-}$. This state will then be referred to as a zero mode. It is the one from which other states can be induced by acting with raising operators $A^{+}$ and $B^{+}$. \par
%
%
Let us first consider the action of $A^{-}$ on functions $f(z,\bar{z})$,
\begin{equation}
  A^{-} f(z,\bar{z}) =0,  \label{eq:E8Afgr1}
\end{equation}
which implies solving the PDE in the variables $z$, $\bar{z}$,
\begin{equation}
  \left(\partial_z + \frac{1}{2} \sqrt{b_1} \bar{z}\right))f(z,\bar{z}) =0, \label{eq:E8Afgr2}
\end{equation}
thus leading to
\begin{equation}
  f(z,\bar{z})= c(\bar{z}) e^{-\frac{1}{2} \sqrt{b_1} z \bar{z} }. \label{eq:E8Afgr3}
\end{equation}
\par
%
%
The other lowering operator $B^{-}$ provides the constraint
\begin{equation}
  B^{-} g(z,\bar{z}) =0, \label{eq:E8Bfgr1}
\end{equation}
which implies the following PDE in the variables $z$, $\bar{z}$,
\begin{equation}
  \left( \partial_{\bar{z}} + \frac{1}{2} \sqrt{b_1} z - \frac{a_3}{2\sqrt{b_1}} \frac{1}{\bar{z}^3} \right) 
  g(z,\bar{z}) =0. \label{eq:E8Bfgr2}
\end{equation}
The solution of the latter is
\begin{equation}
  g= \tilde{c}(z) e^{-\frac{1}{2} \sqrt{b_1} z \bar{z} - \frac{a_3}{4 \sqrt{b_1}} \frac{1}{\bar{z}^2}}.   
  \label{eq:E8Bfgr3}
\end{equation}
\par
%
%
Then the constraints from both lowering operators
\begin{equation}
  A^{-} \psi_0 =B^{-} \psi_0 =0 \label{eq:E8ABf}
\end{equation}
lead to
\begin{equation}
  \psi_0=e^{-\frac{1}{2} \sqrt{b_1} z \bar{z} - \frac{a_3}{4 \sqrt{b_1}} \frac{1}{\bar{z}^2} }. \label{eq:E8gr}
\end{equation}
The action of the Hamiltonian on $\psi_0$ is given by
\begin{equation}
  H\psi_0 =2\sqrt{b_1} \psi_0. \label{eq:E8Hgr}
\end{equation}
\par
%
%
The other states are induced from $\psi_0$ by acting iteratively with raising operators as follows,
\begin{equation}
  \psi_n = (A^{\dagger})^n \psi_0 ,\quad \bar{\psi}_n = (B^{\dagger})^n \psi_0, \label{eq:E8psi}
\end{equation}
and more generally

\begin{equation}
\phi_{m,n}= (A^{\dagger})^m (B^{\dagger})^n \psi_0 ,\quad \phi_{m,0}=\psi_m ,\quad \phi_{0,n}=\bar{\psi}_n, \label{eq:E8psinm}
\end{equation}
with the states $\psi_n$ having the usual interpretation as eigenstates of both the Hamiltonian $H$ and the integral of motion $L_1$,
\begin{equation}
  H \psi_n = 2(n+1) \sqrt{b_1} \psi_n, \label{eq:E8Hpsi}
\end{equation}
\begin{equation}
  L_1 \psi_n =0. \label{eq:E8L1psi}
\end{equation}
\par
%
%
The case of the states $\bar{\psi}_n$ in (\ref{eq:E8psi}) is much less straithforward. We get for the integral $L_1$ that it connects directly the state $\bar{\psi}_n$ to the eigenstates $\psi_n$,
\begin{equation}
   L_1^n \bar{\psi}_n = n! b_1^{\frac{n}{2}} \psi_n . \label{eq:E8L1psib}
\end{equation}
\par
%
%
However, for the integral $R$, for every $\bar{\psi}_n$ there exists some polynomial of $R$ which annihilates the state $\bar{\psi}_n$,
\begin{equation}
  ( R +  \sqrt{b_1} a_3)^{p(n)} \bar{\psi}_n =0, \label{eq:E8Rpsib}
\end{equation}
\begin{equation}
  -1-2n + p(n) + p(n+1) =0, \quad n \geq 4, \label{eq:E8Rpsib2}
\end{equation}
with the first values being
\[ p(1)=1,\quad p(2)=3,\quad p(3)=3,\quad p(4)=5 . \]
We can solve (\ref{eq:E8Rpsib2}), thus getting 
\[ p(n)= \begin{cases} 
                      p(n) =1 &  n=1, \\
		      p(n)=3  &   n=3, \\
		      p(n)=n+1  & n=2,4,6, \ldots, \\
		      p(n)=n-1 &  n=5,7,9, \ldots,
             \end{cases}
\]
where the last two cases can be included together in the formula
\begin{equation}
  p(n)= (-1)^{n-3} [-1+3(-1)^{n-3} +(-1)^{n-3}(n-3) ]. \label{eq:E8Rpsib3}
\end{equation}
As a matter of fact, equations (\ref{eq:E8Rpsib}) and (\ref{eq:E8Rpsib2}) have been checked up to $n=10$ and are conjectured to be valid for higher $n$ values.\par
%
%
Similarly, it has been shown that up to $n=10$, some polynomial of $H$ also annihilates $\bar{\psi}_n$,
\begin{equation}
  \left(H-2(1+n)\sqrt{b_1}\right) \prod_{i=-\frac{n}{2}+1}^{\frac{3n}{2}-1} \left( H +(2+4i)\sqrt{b_1}\right)
  \bar{\psi}_n =0. \label{eq:E8Hpsib}
\end{equation}
\par
%
%
Note that we can also obtain the action of ladder operators on those states, 
\begin{equation}
  A^{-} \psi_n =0 , \quad B^{-} \psi_n  = - n \sqrt{b_1} \psi_{n-1}, \quad A^{-} \bar{\psi}_n = - n \sqrt{b_1} 
  \bar{\psi}_{n-1} . \label{eq:E8ABpsib1}
\end{equation}
Furthermore, from (\ref{eq:E8ABpsib1}) and the commutation relations (\ref{eq:E8lad2}) and (\ref{eq:E8lad3}), we can show that
\begin{equation}
  A^{-} B^{+} \bar{\psi}_n =-(n+1)\sqrt{b_1} \bar{\psi}_n , \quad A^{-} B^{+} \psi_n =-\sqrt{b_1} \psi_n,    
  \label{eq:E8ABpsib2}
\end{equation}
\begin{equation}
  A^{+} B^{-} \psi_n =-n \sqrt{b_1} \psi_n .\label{eq:E8ABpsib3}
\end{equation}
Equations (\ref{eq:E8ABpsib2}) and (\ref{eq:E8ABpsib3}) indicate that in fact those generalized states can have a more direct interpretation in terms of the algebra generated from certain quadratic polynomials of the ladder operators. This can be exploited in order to construct another algebraic structure, namely the dynamical symmetry algebra.\par
%
%
\section{\boldmath Construction of $E_8$ dynamical symmetry algebra}

\setcounter{equation}{0}

Let us now introduce some quadratic polynomials of the ladder operators. More specifically, let us consider the following operators $Q$, $S$, $T$, $U$, $W$, and the function $M$, defined in (\ref{eq:M}), which plays a role in closing the commutation relations,
\begin{equation}
  Q=A^{+}B^{-}, \quad S=B^{+}A^{-}, \quad T=A^{+}A^{-}, \label{eq:E8dynop1}
\end{equation}
\begin{equation}
  U= A^{+}-A^{-}, \quad W=A^{+}+A^{-}, \quad M=[B^{-},B^{+}] . \label{eq:E8dynop2}
\end{equation}
\par
%
%
Those operators satisfy the commutation relations

\begin{equation}
  [Q,S]=M T, \quad [Q,T]=-\sqrt{b_1} T, \quad [Q,U]=-\frac{1}{2}\sqrt{b_1}(U+W),  \label{eq:E8dynal1}
\end{equation}
\begin{equation} 
  [Q,W]= -\frac{1}{2} \sqrt{b_1} (U+W),\quad [Q,M]=-\frac{2}{3b_1 a_3}U^3 W M^2 + 2\sqrt{b_1} M,
   \label{eq:E8dynal2}
\end{equation}
\begin{equation}
  [S,T]= \sqrt{b_1} T , \quad [S,U]= \sqrt{b_1} ( U-W ) , \quad [S,W]=\sqrt{b_1}(-U + W ), \label{eq:E8dynal3}
\end{equation}
\begin{equation}
  [S,M]= - \frac{2}{3 b_1 a_3} U^3 W M^2 -2 \sqrt{b_1} M, \label{eq:E8dynal4}
\end{equation}
\begin{equation}
  [T,U]=[T,W]=[T,M]=0, \label{eq:E8dynal5}
\end{equation}
\begin{equation}
  [U,W]=[U,M]=[W,M]=0, \label{eq:E8dynal6}
\end{equation}
and the following additional relation among those operators
\begin{equation}
 T -\frac{1}{4} W^2 + \frac{1}{4} U^2 =0. \label{eq:E8dynal7}
\end{equation}
\par
%
%
Equations  (\ref{eq:E8dynal1}), (\ref{eq:E8dynal2}), (\ref{eq:E8dynal3}), (\ref{eq:E8dynal4}), (\ref{eq:E8dynal5}), (\ref{eq:E8dynal6}), and (\ref{eq:E8dynal7}) form a closed algebra. On using the differential operator realization of the ladder operators, we can provide the following expressions for the generators of this algebra
\begin{equation}
  Q= A^{+} B^{-} = \partial_z \partial_{\bar{z}} + \left( \frac{1}{2}\sqrt{b_1} z - \frac{a_3}{2\sqrt{b_1}} 
  \frac{1}{\bar{z}^3}\right)\partial_{z} - \frac{1}{2} \sqrt{b_1} \bar{z} \partial_{\bar{z}} 
  -\frac{1}{4} b_1 z \bar{z} + \frac{1}{4} a_3 \frac{1}{\bar{z}^2} + \frac{1}{2} \sqrt{b_1}, 
  \label{eq:E8dynopreal1}
\end{equation}
\begin{equation}
  S= B^{+} A^{-}= \partial_z \partial_{\bar{z}} -\left( \frac{1}{2} \sqrt{b_1} z - \frac{a_3}{2 \sqrt{b_1}} \frac{1}
  {\bar{z}^3}\right)\partial_{z} + \frac{1}{2} \sqrt{b_1} \bar{z} \partial_{\bar{z}} - \frac{1}{4} b_1 z \bar{z} + 
  \frac{1}{4} a_3 \frac{1}{\bar{z}^2} + \frac{1}{2} \sqrt{b_1}, \label{eq:E8dynopreal2}
\end{equation}
\begin{equation}
  T = \partial_z^2 - \frac{1}{4} b_1 \bar{z}^2, \quad U= - \sqrt{b_1} \bar{z}, \quad W = 2\partial_z.
\end{equation}
\par
%
%
The commutation relations of $Q$, $S$, $T$ , $U$, $W$, and $M$ are closed under the commutator, but this is not sufficient to ensure they generate the dynamical symmetry algebra of $E_8$. For such a purpose, it is necessary to check that the Hamiltonian can be written in terms of the generators. On relying on the explicit differential operator (\ref{eq:E8Hr}), it is straightforward to demonstrate that this is indeed the case since
\begin{equation}
  H = - 2 Q - 2 S + 2 \sqrt{b_1}. \label{eq:E8Hdynal}
\end{equation}
We therefore conclude that the algebra generated by $Q$, $S$, $T$, $U$, $W$, and $M$ is the hidden dynamical symmetry algebra of $E_8$.\par
%
%
Another interesting property that can be demonstrated using the explicit differential operator realizations is the connection of the dynamical symmetry algebra generators with the integrals of motion generating the symmetry algebra, $L_1$ and $R$, given in (\ref{eq:E8L1r}) and
\begin{equation}
  R = 4 (z\partial_z - \bar{z} \partial_{\bar{z}} + 1) \left(- \partial_z^2 + \frac{b_1}{4} \bar{z}^2\right) +
  \frac{2a_3}{\bar{z}} \partial_z,
\end{equation}
respectively, namely
\begin{equation}
  L_1 = - T, \label{eq:E8L1dynal}
\end{equation}
\begin{equation}
  R= \frac{4}{\sqrt{b_1}} (S-Q)T -4 T - \frac{1}{6b_1} U M W^3 + \frac{1}{2b_1} U^3 M W. \label{eq:E8Rdynal}
\end{equation}
The topic of connecting different algebraic structures, such as symmetry algebras and dynamical symmetry algebras, for superintegrable systems being relatively unexplored, the present results allow to provide some insight in the subject.\par
%
%
It is also interesting to note the following additional relations involving the integral of motion $R$, 
\begin{equation}
  [R,A^{\pm}]= 4 (A^{\pm})^2 A^{\mp}, \label{eq:E8laddyn1}
\end{equation}
\begin{align}
  [R,B^{+}]&= 4 A^{+}\left(-2 B^{+} A^{-} + A^{+} B^{-} - \frac{1}{\sqrt{b_1}} M A^{+} A^{-}\right)
      + \frac{2}{9 b_1^2 a_3} M^2 U^4 W ( 3 U^2 - W^2) \nonumber \\
  & \quad + \frac{1}{2 \sqrt{b_1}} M\left( -U^3 +3 U^2 W + U W^2 - \frac{1}{3} W^3\right), 
\end{align}
\begin{align}
  [R,B^{-}]&= 4 \left( B^{+} A^{-} -2 A^{+} B^{-} - \frac{1}{\sqrt{b_1}} M A^{+} A^{-}\right) A^{-} 
  + \frac{2}{9 b_1^2 a_3} M^2 U^4 W ( 3 U^2 - W^2) \nonumber \\ 
  & \quad + \frac{1}{2 \sqrt{b_1}} M\left( U^3 +3 U^2 W - U W^2 - \frac{1}{3} W^3\right).
\end{align}
It is quite interesting that relations involving the operator $R$ are closing too, since the latter plays an important role in the closure relation of the quadratic symmetry algebra, in the construction of the Casimir invariant, and in the search of physically relevant representations.\par
%
%
\section{\boldmath New ladder operators for the superintegrable system $E_{10}$ }

\setcounter{equation}{0}

Let us now consider the construction of new ladder operators for the Hamiltonian of $E_{10}$ and demonstrate how they allow to build states which are natural from the integral of motion viewpoint.\par
%
%
\subsection{Construction of ladder operators and generating spectrum algebra}

We have first ladder operators $A^{+}$ and $A^{-}$, defined by
\begin{equation}
  A^{\pm}= \partial_z \mp \frac{1}{2} \sqrt{b_1} \bar{z}, \label{eq:E10Apm}
\end{equation}
and satisfying the relation
\begin{equation}
  [H,A^{\pm}]=\pm 2 \sqrt{b_1} A^{\pm}. \label{eq:E10ApmH}
\end{equation}
We can introduce as well the additional ladder operators $B^{+}$ and $B^{-}$, given by
\begin{equation}
  B^{\pm}=\partial_{\bar{z}} \mp \frac{1}{2} \sqrt{b_1} z \pm \frac{1}{4} \sqrt{b_1} \bar{z}^2 \pm \frac{a_3}
  {2 \sqrt{b_1}}, \label{eq:E10Bpm}
\end{equation}
and such that
\begin{equation}
  [H,B^{\pm}]= \pm 2 \sqrt{b_1} B^{\pm} \pm 2 ( A^{+} - A^{-} ) A^{\pm}. \label{eq:E10BpmH}
\end{equation}
Furthermore, the set of ladder operators fulfil the commutation relations
\begin{align}
  [A^-, A^+] &=0, \quad [B^-, B^+] = A^- - A^+, \label{eq:E10lad1} \\
  [A^{\pm},B^{\pm}]&=0, \quad [A^{\pm}, B^{\mp}] = \pm \sqrt{b_1}. \label{eq:E10lad2} 
\end{align}
\par
%
%
We can close the algebra into a generating spectrum algebra containing both the generators of the quadratic symmetry algebra, with commutation relations (\ref{eq:E10Q1}), (\ref{eq:E10Q2}), (\ref{eq:E10Q3}), and the ladder operators satisfying equations (\ref{eq:E10lad1}), (\ref{eq:E10lad2}). We can indeed write the integrals of motion in terms of the ladder operators as follows,
\begin{equation}
  L_1=-A^{+}A^{-} + \frac{a_3}{12},  \label{eq:E10L1lad}
\end{equation}
\begin{align}
  L_2 &= \frac{1}{b_1} (A^{-} - A^{+})^2 A^{+} A^{-} + \frac{1}{\sqrt{b_1}}[ A^{+}(A^{+}+A^{-})B^{-} - B^{+}
      (A^{+} + A^{-})A^{-}] \nonumber \\
  &\quad + \frac{2a_3}{b_1} A^{+} A^{-} - B^{+} B^{-} - \frac{a_3^2}{4b_1}, \label{eq:E10L2lad}
\end{align}
\begin{equation}
  R= - 2 A^{+}(A^{+} + A^{-})A^{-} - \sqrt{b_1} ( A^{+}B^{-}-B^{+}A^{-} ). \label{eq:E10Rlad}
\end{equation}
From this, we get the following commutation relations of the symmetry algebra generators with the ladder operators,      
\begin{equation}
  [L_1,A^{\pm}]=0, \label{eq:E10symlad1}
\end{equation}
\begin{equation}
  [L_1,B^{\pm}]=\pm \sqrt{b_1} A^{\pm}, \label{eq:E10symlad2}
\end{equation}
\begin{equation}
  [L_2,A^{\pm}]=-(A^{+}+A^{-})A^{\pm} \pm \sqrt{b_1} B^{\pm}, \label{eq:E10symlad3}
\end{equation}
\begin{equation}
  [L_2,B^{+}]= -\frac{2}{\sqrt{b_1}} A^{+}(A^{+}-A^{-})^2 + A^{+} \left( 2B^{+}- B^{-} - \frac{2a_3}
  {\sqrt{b_1}}\right) + B^{+}A^{-}, \label{eq:E10symlad3bis}
\end{equation}
\begin{equation}
  [L_2,B^{-}]= \frac{2}{\sqrt{b_1}} (A^{+}-A^{-})^2 A^{-} +  \left( 2B^{-}- B^{+} + \frac{2a_3}{\sqrt{b_1}}
  \right) A^{-} + A^{+}B^{-}, \label{eq:E10symlad4}
\end{equation}
\begin{equation}
  [R,A^{\pm}]=b_1 A^{\pm}, \label{eq:E10symlad5}
\end{equation}
\begin{equation}
  [R,B^{\pm}]=\pm 3 \sqrt{b_1} ( A^{+} + A^{-} ) A^{\pm} -b_1 B^{\pm}. \label{eq:E10symlad6}
\end{equation}
\par
%
%
\subsection{Construction of states and action of operators}

We will develop the construction of the states in analogy with what was done in section~3.2 by considering first the state annihilated by both lowering operators $A^{-}$ and $B^{-}$,
\begin{equation}
  A^{-}\psi_0 =B^{-} \psi_0 =0.  \label{eq:E10ABgr1}
\end{equation}
This provides the following ground state 
\begin{equation}
  \psi_0 = e^{\frac{\bar{z}}{12 \sqrt{b_1}}(-6b_1z + b_1\bar{z}^2 + 6a_3)}, \label{eq:E10ABgr2}
\end{equation}
from which we build other states by acting iteratively with the creation operators,
\begin{equation}
  \phi_{m,n}=(A^{+})^m (B^{+})^n \psi_0, \label{eq:E10psinm1}
\end{equation}
\begin{equation}
  \psi_m=\phi_{m,0}=(A^{+})^m \psi_0 , \quad \bar{\psi}_n=\phi_{0,n}=(B^{+})^n \psi_0. \label{eq:E10psinm2}
\end{equation}
%
%
The action of the ladder operators on the states (\ref{eq:E10psinm1}) can be calculated in an algebraic way and is given by
\begin{equation}
  A^{+} \phi_{m,n} = \phi_{m+1,n}, \quad A^{-}\phi_{m,n}=-n \sqrt{b_1} \phi_{m,n-1}, \label{eq:E10Apsinm}
\end{equation}
\begin{equation}
  B^{+} \phi_{m,n} =\phi_{m,n+1}, \quad B^{-} \phi_{m,n}=-n \phi_{m+1,n-1} -\frac{n(n-1)}{2}\sqrt{b_1} 
  \phi_{m,n-2} - m \sqrt{b_1} \phi_{m-1,n}. \label{eq:E10Bpsinm}
\end{equation}
%
%
This action can then be used to calculate that of several quadratic expressions 
\begin{equation}
  A^{-} B^{+} \bar{\psi}_n = - (n+1) \sqrt{b_1} \bar{\psi}_n, \label{eq:E10ABpsinm1}
\end{equation}
\begin{equation}
 A^{-} B^{+} \psi_n = - \sqrt{b_1} \psi_n, \label{eq:E10ABpsinm2}
\end{equation}
\begin{equation}
  (A^- B^-)^n \bar{\psi}_n = 0 \quad \text{if $n=1,2,3$},
\end{equation}
\begin{equation}
  (A^{-}B^{-})^{n-1} \bar{\psi}_n =0 \quad \text{if $n \geq 4$}, \label{eq:E10ABpsinm3}
\end{equation}
\begin{equation}
  A^{+} B^{-} \psi_n = - n \sqrt{b_1} \psi_n . \label{eq:E10ABpsinm4}
\end{equation}
These equations indicate that the generalized states have a simpler interpretation in terms of some combinations of ladder operators. This will be exploited in section~6 to construct the dynamical symmetry algebra.\par
%
%
We can as well examine more closely the relations with the Hamiltonian and the integrals of motion, suggesting  connections of the generalized states with the symmetry algebra generators,
\begin{equation}
  H \phi_{m,n} = 2n \phi_{m+2,n-1} + n (n-1) \sqrt{b_1} \phi_{m+1,n-2} + 2 (m+n +1) \sqrt{b_1} \phi_{m,n}, 
  \label{eq:E10Hpsinm}
\end{equation}
\begin{equation}
  L_1 \phi_{m,n} = n \sqrt{b_1} \phi_{m+1,n-1} + \frac{a_3}{12} \phi_{m,n}, \label{eq:E10L1psinm}
\end{equation}
\begin{equation}
  R \phi_{m,n} = 3 n \sqrt{b_1} \phi_{m+2,n-1} - \frac{3}{2} n(n-1) b_1 \phi_{m+1,n-2} + (m-n)b_1 \phi_{m,n}. 
  \label{eq:E10Rpsinm}
\end{equation}
\par
%
%
Furthermore, we can obtain several formulas expressing the action of polynomials of these operators on $\bar{\psi}_n$. To start with, the relation 
\begin{align}
  &\prod_{i=0}^{p-1} [ R + (n-3i)b_1] \bar{\psi}_n = \sum_{q=0}^{p} 3^p \frac{n!}{(n-p-q)!} 
      \left(-\frac{1}{2}\right)^q \begin{pmatrix} p \\ q \end{pmatrix} b_1^{\frac{p+q}{2}} \phi_{2p-q,n-p-q}, 
      \nonumber \\
  & \qquad p=1,2,\ldots, n, \label{eq:E10Rpolpsinm1}
\end{align}
can be easily proved by induction over $p$. In particular, for $p=n$, it writes
\begin{equation}
  \prod_{i=0}^{n-1} [ R+(n-3i)b_1]\bar{\psi}_n = 3^n n! b_1^{\frac{n}{2}} \psi_{2n}, \label{eq:E10Rpolpsinm2}
\end{equation}
so that
\begin{equation}
  \prod_{i=0}^n [ R+(n-3i)b_1]\bar{\psi}_n = 0.
\end{equation}
\par
%
%
{}For the integral $L_1$, the relation
\begin{equation}
  \left( L_1 - \frac{a_3}{12}\right)^p \bar{\psi}_n = \frac{n!}{(n-p)!} b_1^{\frac{p}{2}} \phi_{p,n-p}, \quad 
  p=1,2,\ldots,n, \label{eq:E10L1polpsinm1}
\end{equation}
can also be proved by induction over $p$. As a special case, for $p=n$, we get
\begin{equation}
 \left( L_1- \frac{a_3}{12}\right)^n \bar{\psi}_n = n! b_1^{\frac{n}{2}} \psi_n. \label{eq:E10L1polpsinm2}
\end{equation}
\par
%
%
Similarly, it has been shown up to $n=20$ and 21 that some polynomial of $H$ also annihilates $\bar{\psi}_n$,
\begin{equation}
  \prod_{i=i_{\rm min}}^{2n+1} \bigl( H-2i \sqrt{b_1}\bigr) \bar{\psi}_n = 0,  \label{eq:E10Hpolpsinm1}                                              
\end{equation}
where $i_{\rm min} = n+1-m$ for $n=2m$ or $2m+1$.\par
%
%
The results derived above show the interest of considering the states generated by the complete set of ladder operators. Those states would not be obtained in the usual approach based on Hermitian superintegrable systems.%
%
\section{ Construction of $E_{10}$ dynamical symmetry algebra}

\setcounter{equation}{0}

The key in constructing the hidden dynamical symmetry algebra consists in considering a bilinear transformation
from the complete set of ladder operators. Let us introduce the operators $Q$ , $S$, $T$, $U$, $W$, defined by
\begin{equation}
  Q= A^{+} B^{-}, \quad  S =B^{+} A^{-}, \quad T=A^{+}A^{-}, \quad U=A^{+}-A^{-}, \quad W=A^{+} + A^{-}.
   \label{eq:E10dynop5}
\end{equation}
\par
%
%
These new operators satisfy the following quadratic algebra,
\begin{equation}
  [Q,S]= -T U , \quad [Q,T]=-\sqrt{b_1} T, \quad [S,T]=\sqrt{b_1}T, \label{eq:E10dynal1}
\end{equation}
\begin{equation}
  [Q,U]=-\frac{\sqrt{b_1}}{2} U - \frac{\sqrt{b_1}}{2} W, \quad [S,U]=\frac{\sqrt{b_1}}{2} U - 
  \frac{\sqrt{b_1}}{2} W, \label{eq:E10dynal2}
\end{equation}
\begin{equation}
  [Q,W]=-\frac{\sqrt{b_1}}{2} U - \frac{\sqrt{b_1}}{2} W, \quad [S,W]=-\frac{\sqrt{b_1}}{2} U +  
  \frac{\sqrt{b_1}}{2} W, \label{eq:E10dynal3}
\end{equation}
\begin{equation}
  [T,U]=[T,W]=[U,W]=0. \label{eq:E10dynal4}
\end{equation}
The structure of such an algebra is rather reminiscent of that of a non-semisimple Lie algebra, apart from the fact that there is an additional quadratic relation among the generators. \par
%
%
In order to demonstrate this is the dynamical symmetry algebra, we need to establish that the Hamiltonian can be expressed in terms of its generators. On using the explicit expressions in terms of the differential operators in $z$ and $\bar{z}$, we indeed find
\begin{equation}
  H=-2Q -2S + 2 \sqrt{b_1}. \label{eq:E10Hdynop}
\end{equation}
The integrals of motion $L_1$ and $R$ can also be written in terms of the generators since
\begin{equation}
  L_1 =-T + \frac{a_3}{12}, \label{eq:E10L1dynop}
\end{equation}
\begin{equation}
  R=-2 W T - \sqrt{b_1} ( Q-S ). \label{eq:E10Rdynop}
\end{equation}
Equations (\ref{eq:E10Hdynop}), (\ref{eq:E10L1dynop}), and (\ref{eq:E10Rdynop}) show as well the close connection existing between the dynamical symmetry algebra and the symmetry algebra. Here the integrals of motion are completely determined by the generators of the dynamical symmetry algebra, although the latter lies beyond usual Lie algebras.\par
%
%
\section{Conclusion}

We have considered two models from the classification of superintegrable systems on two-dimensional conformally flat spaces. We re-examined those models with a new approach that was introduced in the context of pseudo-Hermitian harmonic and anharmonic oscillators. In one case, for the model $E_{10}$, the ladder operators satisfy a Lie algebra, however, in the other case $E_{8}$, the commutation relations are nonlinear. For both Hamiltonians, we also presented the generating spectrum algebra, obtained by combining the ladder operators with the symmetry algebra, the latter taking the form of a quadratic algebra in both cases.\par
%
%
We then used the complete set of ladder operators to construct the states and we also presented several formulas for the action of the operators on the latter. In both cases, we demonstrated that there exist some states $\bar{\psi}_n$ that can be annihilated by polynomials of the integral $R$ or polynomials of the Hamiltonian $H$, as well as being connected to the eigenstates $\psi_n$ via the action of $L_1$. Those generalized states are not obtained in the usual approach to superintegrable systems. We also constructed the hidden dynamical symmetry algebra and showed how its generators are not only connected to the Hamiltonian, but also to the integrals of motion. This allowed us to introduce a new approach to superintegrable systems constructed on complex spaces and to display properties that are usually considered in the context of pseudo-Hermitian systems.\par
%
%
The possibility of relating some complex potentials from the 58 two-dimensional superintegrable systems through projective equivalence has been discussed \cite{vol21,vol18} and different types of normal form, Liouville, complex Liouville, and Dini associated with Jordan blocks have been examined. The methods developed in this paper and the related methods will have applications to those models with complex Liouville and Dini form.\par
%
%
The search of three-dimensional superintegrable systems is difficult \cite{kal06b,kal06c}, in particular semi-degenerate and degenerate ones \cite{ruiz17}. Other ideas on using systems of algebraic equations \cite{kress14,kress19} have been proposed, as well as relying on different underlying algebraic structures \cite{ford20}. As there is  some progress on the classification of three-dimensional superintegrable systems and beyond, the methods presented in this paper are likely to be of interest for many of those models. This points out the wider applicability of the algebraic methods used here.\par
%
%
Superintegrable systems have several applications. Recently, it was demonstrated how two-dimensional superintegrable systems can be used to describe systems with three-body interaction \cite{tur21}. This points out again the possibility of applying the new methods presented here to other models to obtain generalized states.\par
%
%
\section*{Acknowledgments}

IM was supported by Australian Research Council Fellowship FT180100099. CQ was supported by the Fonds de la Recherche Scientifique - FNRS under Grant Number 4.45.10.08.\par
%
%


\newpage

\begin{thebibliography}{99}

\bibitem{kal05a}
Kalnins E G, Kress J M and Miller W Jr 2005 
Second order superintegrable systems in conformally flat spaces. I. Two-dimensional classical structure theory 
{\sl J.\ Math.\ Phys.} {\bf 46} 053509 

\bibitem{kal05b}
Kalnins E G, Kress J M and Miller W Jr 2005
Second order superintegrable systems in conformally flat spaces. II. The classical two-dimensional St\"ackel transform 
{\sl J.\ Math.\ Phys.} {\bf 46} 053510 

\bibitem{kal05c}
Kalnins E G, Kress J M and Miller W Jr 2005
Second order superintegrable systems in conformally flat spaces. III. Three-dimensional classical structure theory
{\sl J.\ Math.\ Phys.} {\bf 46} 103507

\bibitem{kal06a}
Kalnins E G, Miller W Jr and Pogosyan G S 2006 
Exact and quasiexact solvability of second-order superintegrable quantum systems. I. Euclidean space preliminaries {\sl J.\ Math.\ Phys.} {\bf 47} 033502

\bibitem{kal06b}
Kalnins E G, Kress J M and Miller W Jr 2006
Second order superintegrable systems in conformally flat spaces. IV. The classical 3D St\"ackel transform and 3D classification theory
{\sl J.\ Math.\ Phys.} {\bf 47} 043514

\bibitem{kal06c}
Kalnins E G, Kress J M and Miller W Jr 2006
Second-order superintegrable systems in conformally flat spaces. V. Two- and three-dimensional quantum systems {\sl J.\  Math.\ Phys.} {\bf 47} 093501

\bibitem{kal07}
Kalnins E G, Miller W Jr and Pogosyan G S 2007 
Exact and quasiexact solvability of second order superintegrable quantum systems. II. Relation to separation of variables
{\sl J.\ Math.\ Phys.} {\bf 48} 023503

\bibitem{post08}
Kalnins E G, Miller W Jr and Post S 2008
Models for quadratic algebras associated with second order superintegrable systems in 2D
{\sl SIGMA} {\bf 4} 008 

\bibitem{post09}       
Kalnins E G, Miller W Jr and Post S 2009
Models of quadratic quantum algebras and their relation to classical superintegrable systems
{\sl Phys.\ Atom.\ Nuclei} {\bf 72} 801

\bibitem{post13}  
Kalnins E G, Miller W Jr and Post S 2013
Contractions of 2D 2nd order quantum superintegrable systems and the Askey scheme for hypergeometric orthogonal polynomials
{\sl SIGMA} {\bf 9} 057

\bibitem{mil14a}  
Miller W Jr 2014
The theory of contractions of 2D 2nd order quantum superintegrable systems and its relation to the Askey scheme for hypergeometric orthogonal polynomials
{\sl J. Phys.: Conf.\ Ser.} {\bf 512} 012012

\bibitem{mil14b}
Kalnins E G and Miller W Jr 2014
Quadratic algebra contractions and second order superintegrable systems
{\sl Analysis and Applications} {\bf 12} 583

\bibitem{bender90}
Bender C M, Boettcher S 1990
Real spectra in non-Hermitian Hamiltonians having $\cal PT$ symmetry
{\sl Phys.\ Rev.\ Lett.} {\bf 80} 5243

\bibitem{bender05}
Bender C M 2005
Introduction to $ \cal PT$-symmetric quantum theory
{\sl Contemp.\ Phys.} {\bf 46} 277

\bibitem{bender07}
Bender C M 2007
Making sense of non-Hermitian Hamiltonians
{\sl Rep.\ Prog.\ Phys.} {\bf70} 947

\bibitem{mosta02a}
Mostafazadeh A 2002
Pseudo-Hermiticity versus PT symmetry: The necessary condition for the reality of the spectrum of a non-Hermitian Hamiltonian
{\sl J.\ Math.\ Phys.} {\bf 43} 205

\bibitem{mosta10}
Mostafazadeh A 2010
Pseudo-Hermitian representation of quantum mechanics
{\sl Int.\ J.\ Geom.\ Meth.\ Mod.\ Phys.} {\bf 7} 1191

\bibitem{bender01}
Bender C M, Dunne G V, Meisinger P N and \c Sim\c sek M 2001
Quantum complex H\'enon-Heiles potential
{\sl Phys.\ Lett.} A {\bf 281} 311

\bibitem{nana}
Nanayakhara A 2002
Real eigenspectra in non-Hermitian multidimensional Hamiltonians
{\sl Phys.\ Lett.} A {\bf 304} 67

\bibitem{mosta02b}
Mostafazadeh A 2002
Pseudo-Hermiticity for a class of nondiagonalizable Hamiltonians
{\sl J.\ Math.\ Phys.} {\bf 43} 6343

\bibitem{mosta03}
Mostafazadeh A 2003
Erratum: Pseudo-Hermiticity for a class of nondiagonalizable Hamiltonians [J.\ Math.\ Phys.\ 43, 6343 (2002)]
{\sl J.\ Math.\ Phys.} {\bf 44} 943

\bibitem{ioffe}
Ioffe M V, Cannata F and Nishnianidze D N 2006
Exactly solvable two-dimensional complex model with a real spectrum
{\sl Theor.\ Math.\ Phys.} {\bf 148} 960

\bibitem{cannata10}
Cannata F, Ioffe M V and Nishnianidze D N 2010
Exactly solvable nonseparable and nondiagonalizable two-dimensional model with quadratic complex interaction
{\sl J.\ Math.\ Phys.} {\bf 51} 022108

\bibitem{cannata12}
Cannata F, Ioffe M V and Nishnianidze D N 2012
Equidistance of the complex two-dimensional anharmonic oscillator spectrum: the exact solution
{\sl J.\ Phys.\ A: Math.\ Theor.} {\bf 45} 295303

\bibitem{barda}
Bardavelidze M S, Cannata F, Ioffe M V and Nishnianidze D N 2013
Three-dimensional shape invariant non-separable model with equidistant spectrum
{\sl J.\ Math.\ Phys.} {\bf 54} 012107

\bibitem{marquette22a}
Marquette I and Quesne C 2022
Ladder operators and hidden algebras for shape invariant nonseparable and nondiagonalizable models with quadratic complex interaction. I. Two-dimensional model
{\sl SIGMA} {\bf 18} 004

\bibitem{marquette22b}
Marquette I and Quesne C 2022 
Ladder operators and hidden algebras for shape invariant nonseparable and nondiagonalizable models with quadratic complex interaction. II. Three-dimensional model
{\sl SIGMA} {\bf 18} 005

\bibitem{marquette21a}
Marquette I and Quesne C 2021 
Algebraic construction of associated functions of nondiagonalizable models with anharmonic oscillator complex interaction
(arXiv:2111.01617[quant-ph])

\bibitem{das99}
Bonatsos D and Daskaloyannis C 1999
Quantum groups and their applications in nuclear physics,
{\sl Prog.\ Part.\ Nucl.\ Phys.\ } {\bf 43} 537 

\bibitem{ver08}
Verrier P E and Evans N W 2008
Superintegrability of the caged anisotropic oscillator {\sl J.\ Math.\ Phys.\ } {\bf 49} 092902 

\bibitem{rod08}
Rodriguez M A, Tempesta P and Winternitz P 2008
Reduction of superintegrable systems: The anisotropic harmonic oscillator, {\sl Phys.\ Rev.\ E} {\bf 78} 046608 

\bibitem{kal10}
Kalnins E G, Miller Jr W and Post S 2010 
Superintegrability and higher order integrals for quantum systems, {\sl J.\ Phys.\ A: Math.\ theor.\ } {\bf 43} 035202 

\bibitem{mar10}
Marquette I 2010 
Superintegrability and higher order polynomial algebras, 
{\sl J.\ Phys.\ A: Math.\ Theor.\ } {\bf 43} 135203 

\bibitem{vol21}
Vollmer A 2021 
St\"ackel equivalence of non-degenerate superintegrable systems and invariant quadrics
{\sl SIGMA} {\bf 17} 015

\bibitem{vol18}
Vollmer A 2020 
Projectively equivalent 2-dimensional superintegrable systems with projective symmetries
{\sl J.\ Phys.\ A: Math.\ Theor.} {\bf 53} 095202

\bibitem{ruiz17}
Escobar-Ruiz M A and Miller W Jr 2017 
Toward a classification of semidegenerate 3D superintegrable systems
{\sl J. Phys. A: Math.\ Theor.} {\bf 50} 095203

\bibitem{kress14}
Capel J J and Kress J M 2014
Invariant classification of second-order conformally flat superintegrable systems
{\sl J. Phys. A: Math.\ Theor.} {\bf 47} 495202

\bibitem{kress19}
Kress J M and Sch\"obel K 2019
An algebraic geometric classification of superintegrable systems in the Euclidean plane 
{\sl J.\ Pure Appl.\ Algebra} {\bf 223} 1728

\bibitem{ford20}
Fordy A P and Huang Q 2020 
Superintegrable systems on 3 dimensional conformally flat spaces
{\sl J.\ Geom.\ Phys.} {\bf 153} 103687

\bibitem{tur21}
Turbiner A V, Miller W Jr and Escobar-Ruiz M A 2021
From two-dimensional (super-integrable) quantum dynamics to (super-integrable) three-body dynamics
{\sl J.\ Phys.\ A: Math.\ Theor.} {\bf 54} 015204

\end{thebibliography}
\end{document}